%% file: main.tex
\documentclass[sigconf]{acmart}
\pdfoutput=1
\AtBeginDocument{%
  \providecommand\BibTeX{{%
    \normalfont B\kern-0.5em{\scshape i\kern-0.25em b}\kern-0.8em\TeX}}}
\usepackage{multirow}
\usepackage{xspace}
\usepackage{soul}
\usepackage[noabbrev,capitalize]{cleveref}
\usepackage{enumitem}
\usepackage{adjustbox}
\usepackage{pgfplots}
\usepackage{makecell}
\usepackage{multirow}
\usepackage{pgfplots}
\usepackage{adjustbox}
\usepackage{tikz}
\usepackage{subcaption}
\usepackage{bbm}
\usepgfplotslibrary{colorbrewer}
\usetikzlibrary{calc,shadings,patterns}
\usetikzlibrary{fit}

\pgfplotsset{compat=1.18}

\newenvironment{customlegend}[1][]{%
\begingroup
\csname pgfplots@init@cleared@structures\endcsname
\pgfplotsset{#1}%
}{%
\csname pgfplots@createlegend\endcsname
\endgroup
}%
\def\addlegendimage{\csname pgfplots@addlegendimage\endcsname}

\renewcommand\footnotetextcopyrightpermission[1]{}
\author{Daniele Malitesta}
\authornote{Work done during a research visit at the University of Edinburgh.}
\email{daniele.malitesta@centralesupelec.fr}
\affiliation{\institution{Université Paris-Saclay, CentraleSupélec, Inria}
\country{Gif-sur-Yvette, France}
 }

\author{Alberto Carlo Maria Mancino}
\authornotemark[1]
\email{alberto.mancino@poliba.it}
\affiliation{\institution{Politecnico di Bari}
\country{Bari, Italy}
 }

\author{Pasquale Minervini}
\email{p.minervini@{ed.ac.uk, miniml.ai}}
\affiliation{\institution{University of Edinburgh \& Miniml.AI}
\country{Edinburgh, United Kingdom}
 }

\author{Tommaso Di Noia}
\email{tommaso.dinoia@poliba.it}
\affiliation{\institution{Politecnico di Bari}
\country{Bari, Italy}
 }

\setcopyright{acmcopyright}
\copyrightyear{2018}
\acmYear{2018}
\acmDOI{10.1145/1122445.1122456}

\acmConference[Woodstock '18]{Woodstock '18: ACM Symposium on Neural
  Gaze Detection}{June 03--05, 2018}{Woodstock, NY}
\acmBooktitle{Woodstock '18: ACM Symposium on Neural Gaze Detection,
  June 03--05, 2018, Woodstock, NY}
\acmPrice{15.00}
\acmISBN{978-1-4503-XXXX-X/18/06}

\usepackage{amsmath} 
\usepackage{bm}

\usepackage{xspace}
\newcommand*{\eg}{e.g.\@\xspace}
\newcommand*{\ie}{i.e.\@\xspace}

\newcommand{\norm}[1]{\left\lVert#1\right\rVert}

\begin{abstract}
\input{src/abstract}
\end{abstract}

\begin{document}


\title[Dot Product is All You Need]{Dot Product is All You Need: Bridging the Gap Between Item Recommendation and Link Prediction}





\keywords{Item Recommendation, Link Prediction, Dot Product}


\maketitle

\input{src/introduction}
\input{src/background}
\input{src/proposed}

\input{src/experiments}
\input{src/results_discussion}
\input{src/conclusion}

\section*{Acknowledgements}
Pasquale Minervini was partially funded by ELIAI (The Edinburgh Laboratory for Integrated Artificial Intelligence), EPSRC (grant no. EP/W002876/1), an industry grant from Cisco, and a donation from Accenture LLP.
This work was carried out while Alberto C. M. Mancino was enrolled in the Italian National Doctorate on Artificial Intelligence run by the Sapienza University of Rome in collaboration with the Polytechnic University of Bari.
This work was supported by the Edinburgh International Data Facility (EIDF) and the Data-Driven Innovation Programme at the University of Edinburgh.

\bibliographystyle{ACM-Reference-Format}
\bibliography{bibliography,refs}

\end{document}

%% file: src/abstract.tex
Item recommendation (the task of predicting if a user may interact with new items from the catalogue in a recommendation system) and link prediction (the task of identifying missing links in a knowledge graph) have long been regarded as distinct problems.
In this work, we show that the item recommendation problem can be seen as an instance of the link prediction problem, where entities in the graph represent users and items, and the task consists of predicting missing instances of the relation type <<interactsWith>>.
In a preliminary attempt to  demonstrate the assumption, we decide to test three popular factorisation-based link prediction models on the item recommendation task, showing that their predictive accuracy is competitive with ten state-of-the-art recommendation models. The purpose is to show how the former may be seamlessly and effectively applied to the recommendation task without any specific modification to their architectures. Finally, while beginning to unveil the key reasons behind the recommendation performance of the selected link prediction models, we explore different settings for their hyper-parameter values, paving the way for future directions.

%% file: src/introduction.tex
\section{Introduction and Motivations}

As of today, item recommendation and link prediction are considered and treated as separate tasks. On the one side, item recommendation~\cite{DBLP:reference/rsh/2011} is about retrieving the list of items that may match the preferences of any user in the system (e.g., an e-commerce platform). On the other side, link prediction~\cite{DBLP:conf/cikm/Liben-NowellK03} aims to fill in the blanks in any knowledge graph connecting entities through predicates (e.g., a knowledge graph describing structured information extracted from the Web). 

Nevertheless, there exist at least two aspects that may bring together the two tasks and model families. Indeed, it is trivial to re-cast the item recommendation problem under the link prediction perspective, where users and items in the system are well-represented by entities in a knowledge graph with a single-type predicate, namely, <<interactsWith>>. Furthermore, the literature outlines that both families of models in recommendation and link prediction share common bases, especially in the way they learn the representations of entities (users and items). In this respect, factorisation-based techniques constitute the leading approaches, assuming that entities (users and items) are mapped to embeddings in the latent space, and the interaction score is well-estimated through their \textbf{dot product}. 

In light of the above, and to the best of our knowledge, this paper comes as one of the first (preliminary) attempts in the literature to bridge the gap between item recommendation and link prediction. Thus, we take three popular factorisation-based models for link prediction (i.e., DistMult~\cite{DBLP:journals/corr/YangYHGD14a}, CP~\cite{DBLP:conf/icml/LacroixUO18}, and ComplEx~\cite{DBLP:conf/icml/TrouillonWRGB16}), and test their performance on three widely-adopted recommendation datasets (i.e., Gowalla~\cite{DBLP:conf/www/LiangCMB16}, Yelp 2018~\cite{DBLP:conf/sigir/0001DWLZ020}, and Amazon Book~\cite{DBLP:conf/www/HeM16}). Then, we compare their performance to ten state-of-the-art recommendation models (i.e., MostPop, Random, UserkNN~\cite{DBLP:conf/cscw/ResnickISBR94}, ItemkNN~\cite{DBLP:conf/www/SarwarKKR01}, NGCF~\cite{DBLP:conf/sigir/Wang0WFC19}, DGCF~\cite{DBLP:conf/sigir/WangJZ0XC20}, LightGCN~\cite{DBLP:conf/sigir/0001DWLZ020}, SGL~\cite{DBLP:conf/sigir/WuWF0CLX21}, UltraGCN~\cite{DBLP:conf/cikm/MaoZXLWH21}, and GFCF~\cite{DBLP:conf/cikm/ShenWZSZLL21}) which largely build on the factorisation technique. Note that, for the sake of this explorative work, we do not apply any modification to the link prediction models; indeed, our assumption (based upon the above) is that such models may seamlessly and effectively work in the item recommendation setting. 

Preliminary results show that link prediction approaches can lead to comparable performance to the tested recommendation systems. Finally, to further validate our rationale, we begin to explore different settings for the hyper-parameter values of the selected link prediction models. These additional outcomes indicate that there might exist room for future development and refining of these solutions when applied to the task of personalised recommendation.

The rest of the paper is structured as follows. First, in~\Cref{sec:background}, we provide formal definitions and formulations for the two tasks and model families. Second, in~\Cref{sec:cast}, we outline how the item recommendation problem may be re-cast under the link prediction task, outlining the main research questions for this work. Then, in~\Cref{sec:experiments}, we present the experimental settings for the conducted analysis. Finally, in~\Cref{sec:results}, we present the results and discuss them, paving the way for future directions of this work in~\Cref{sec:conclusions}.

%% file: src/background.tex
\section{Background}
\label{sec:background}
This section provides useful background notions for link prediction and item recommendation. Specifically, we outline the main definitions for the two considered tasks, along with the formulations of popular models in terms of scoring, loss, and regularisation functions.

\subsection{Link prediction}
\subsubsection{Link prediction in knowledge graphs}
A Knowledge Graph $\mathcal{G} \subseteq \mathcal{E} \times \mathcal{R} \times \mathcal{E}$ contains a set of subject-predicate-object $\langle s, p, o \rangle$ triples, where each triple represents a relationship of type $p \in \mathcal{R}$ between the subject $s \in \mathcal{E}$ and the object $o \in \mathcal{E}$ of the triple.
Here, $\mathcal{E}$ and $\mathcal{R}$ denote the set of all entities and relation types, respectively.
However, many real-world knowledge graphs are largely incomplete~\cite{DBLP:conf/aaai/DettmersMS018, DBLP:journals/semweb/FarberBMR18, DBLP:journals/pieee/Nickel0TG16, DBLP:journals/ivs/DestandauF21} -- link prediction focuses of the problem of identifying missing links in (possibly very large) knowledge graphs.
More formally, given an incomplete graph $\mathcal{G}^{-} \subset \mathcal{G}$, where $\mathcal{G}$ denotes a complete graph, the task consists of identifying the triples $\langle s, p, o \rangle$ triples such that $\langle s, p, o \rangle \not\in \mathcal{G}^{-}$ and $\langle s, p, o \rangle \in \mathcal{G}$.
%
%
%

\subsubsection{Link predictors}
A \emph{link predictor} is a differentiable model where entities in $\mathcal{E}$ and relation types in $\mathcal{R}$ are represented in a continuous embedding space, and the likelihood of a link between two entities is a function of their representations.
More formally, link predictors are defined by a parametric \emph{scoring function} $\phi_{\theta} : \mathcal{E} \times \mathcal{R} \times \mathcal{E} \mapsto \mathbb{R}$, with parameters $\theta$ that, given a triple $\langle s, p, o \rangle$, produces the likelihood that entities $s$ and $o$ are related by the relationship $p$.
\subsubsection{Scoring functions}
Link prediction models can be characterised by their scoring function $\phi_{\theta}$.
For example, in TransE~\cite{DBLP:conf/nips/BordesUGWY13}, the score of a triple $\langle s, p, o \rangle$ is given by:
\begin{equation}
\phi^{\text{TransE}}_{\theta}(s, p, o) = - \norm{\mathbf{s} + \mathbf{p} - \mathbf{o}}_{2}, 
\end{equation}
where $\mathbf{s}, \mathbf{p}, \mathbf{o} \in \mathbb{R}^{k}$ denote the embedding representations of $s$, $p$, and $o$, respectively.
In DistMult~\cite{DBLP:journals/corr/YangYHGD14a}, the scoring function is defined as follows:
\begin{equation}
\phi^{\text{DistMult}}_{\theta}(s, p, o) = \langle \mathbf{s}, \mathbf{p}, \mathbf{o} \rangle = \sum_{i=1}^{k} \mathbf{s}_{i} \mathbf{p}_{i} \mathbf{o}_{i},
\end{equation}
where $\langle{}\cdot{}, {}\cdot{}, {}\cdot{} \rangle$ denotes the tri-linear dot product.
Canonical Tensor Decomposition (i.e., CP)~\cite{DBLP:conf/icml/LacroixUO18} is similar to DistMult, with the difference that each entity $x$ has two representations, $\mathbf{x}_{s} \in \mathbb{R}^{k}$ and $\mathbf{x}_{o} \in \mathbb{R}^{k}$, depending on whether it is being used as a subject or object:
\begin{equation}
\phi^{\text{CP}}_{\theta}(s, p, o) = \langle \mathbf{s}_{s}, \mathbf{p}, \mathbf{o}_{o} \rangle.
\end{equation}
In RESCAL~\cite{DBLP:conf/icml/NickelTK11}, the scoring function is a bilinear model given by:
\begin{equation}
\phi^{\text{RESCAL}}_{\theta}(s, p, o) = \mathbf{s}^{\top} \mathbf{P} \mathbf{o},
\end{equation}
where $\mathbf{s}, \mathbf{o} \in \mathbb{R}^{k}$ is the embedding representation of $s$ and $p$, and $\mathbf{P} \in \mathbb{R}^{k \times k}$ is the representation of $p$.
Note that DistMult is equivalent to RESCAL if $\mathbf{P}$ is constrained to be diagonal.
Another variation of this model is ComplEx~\cite{DBLP:conf/icml/TrouillonWRGB16}, where the embedding representations of $s$, $p$, and $o$ are complex vectors -- \ie $\mathbf{s}, \mathbf{p}, \mathbf{o} \in \mathbb{C}^{k}$ -- and the scoring function is given by:
\begin{equation}
\phi^{\text{ComplEx}}_{\theta}(s, p, o) = \Re(\langle \mathbf{s}, \mathbf{p}, \overline{\mathbf{o}} \rangle),
\end{equation}
where $\Re(\mathbf{x})$ represents the real part of $\mathbf{x}$, and $\overline{\mathbf{x}}$ denotes the complex conjugate of $\mathbf{x}$.
In TuckER~\cite{DBLP:conf/emnlp/BalazevicAH19}, the scoring function is defined as follows:
\begin{equation}
\phi^{\text{TuckER}}_{\theta}(s, p, o) = \mathbf{W} \times_{1} \mathbf{s} \times_{2} \mathbf{p} \times_{3} \mathbf{o},
\end{equation}
where $\mathbf{W} \in \mathbb{R}^{k_{s} \times k_{p} \times k_{o}}$ is a three-way tensor of parameters, and $\mathbf{s} \in \mathbb{R}^{k_{s}}$, $\mathbf{p} \in \mathbb{R}^{k_{p}}$, and $\mathbf{o} \in \mathbb{R}^{k_{o}}$ are the embedding representations of $s$, $p$, and $o$.
In this work, we mainly focus on DistMult, CP, and ComplEx due to their effectiveness on several link prediction benchmarks~\cite{DBLP:conf/iclr/RuffinelliBG20, DBLP:journals/corr/abs-2005-00804}.
\subsubsection{Training objectives}
Another dimension for characterising link predictors is their \emph{training objective}.
Early link prediction models such as RESCAL and CP were trained to minimise the reconstruction error of the whole adjacency tensor~\cite{DBLP:conf/icml/NickelTK11, DBLP:conf/acssc/VervlietDL16, DBLP:journals/jmlr/KossaifiPAP19}.
To scale to larger Knowledge Graphs, subsequent approaches such as \citet{DBLP:conf/nips/BordesUGWY13} and \citet{DBLP:journals/corr/YangYHGD14a} simplified the training objective by using \emph{negative sampling}: for each training triple, a corruption process generates a batch of negative examples by corrupting the subject and object of the triple, and the model is trained by increasing the score of the training triple while decreasing the score of its corruptions.
More formally, the loss is given by $\mathcal{L}(\mathcal{G}) = \sum_{\langle s, p, o \rangle} \ell(s, p, o)$ with:
\begin{equation}
\begin{aligned}
\ell(s, p, o) = \sum_{\langle \hat{s}, p, \hat{o} \rangle \in \mathcal{N}(s, p, o)} \left[ \gamma - \phi(s, p, o) + \phi(\hat{s}, p, \hat{o}) \right]_{+},
\end{aligned}
\end{equation}
\noindent where $\mathcal{N}(s, p, o) = \{ \langle \hat{s}, p, o \rangle \mid s \neq \hat{s} \} \cup \{ \langle s, p, \hat{o} \rangle \mid o \neq \hat{o} \}$ denotes the set of triples obtained by corrupting the training triple $\langle s, p, o \rangle$.
This approach was later extended by \citet{DBLP:conf/aaai/DettmersMS018} where, given a subject $s$ and a predicate $p$, the task of predicting the correct objects is cast as a $|\mathcal{E}|$-dimensional multi-label classification task, where each label corresponds to a distinct object and multiple labels can be assigned to the $(s, p)$ pair.
This training objective is referred to as KvsAll by \citet{DBLP:conf/iclr/RuffinelliBG20} and can be formalised as $\mathcal{L}(\mathcal{G}) = \sum_{e_{1} \in \mathcal{E}} \sum_{p \in \mathcal{R}} \ell(e_{1}, p)$, with:
\begin{equation}
\begin{aligned}
\ell(e_{1}, p) = & \sum_{e_{2} \in \mathcal{E}} \text{BCE}(e_{2}, p, e_{1}) + \text{BCE}(e_{1}, p, e_{2}), \\
\end{aligned}
\end{equation}
\noindent with $\text{BCE}(s, p, o) = - (y \log p + (1 - y) \log (1 - p))$, $p = \sigma(\phi(s, p, o))$, and $y = \mathbbm{1}\left[ \langle s, p, o \rangle \in \mathcal{G} \right]$.
Another extension was proposed by \citet{DBLP:conf/icml/LacroixUO18} where, given a subject $s$ and an object $p$, the task of predicting the correct object $o$ in the training triple is cast as a $|\mathcal{E}|$-dimensional multi-class classification task, where each class corresponds to a distinct object and only one class can be assigned to the $(s, p)$ pair; this is referred to as 1vsAll by \citet{DBLP:conf/iclr/RuffinelliBG20}, and defined as $\mathcal{L}(\mathcal{G}) = \sum_{\langle s, p, o \rangle} \ell_{\text{s}}(s, p, o) + \ell_{\text{o}}(s, p, o)$, with:
\begin{equation}
\begin{aligned}
\ell_{\text{s}}(s, p, o) = & - \phi(s, p, o) + \log\left[ \sum_{\hat{s}} \exp\left( \phi(\hat{s}, p, o) \right) \right], \\
\ell_{\text{o}}(s, p, o) = & - \phi(s, p, o) + \log\left[ \sum_{\hat{o}} \exp\left( \phi(s, p, \hat{o}) \right) \right].
\end{aligned}
\end{equation}
\subsubsection{Regularisers}
As noted by \citet{DBLP:conf/nips/BordesUGWY13}, imposing regularisation terms on the learned entity and relation representations prevents the training process from trivially optimising the training objective by increasing the embedding norms.
Early works such as \citet{DBLP:conf/nips/BordesUGWY13,DBLP:conf/aaai/BordesWCB11,DBLP:journals/corr/abs-1301-3485,DBLP:conf/nips/JenattonRBO12} proposed constraining the embedding norms.
More recently, \citet{DBLP:journals/corr/YangYHGD14a,DBLP:conf/icml/TrouillonWRGB16} proposed adding a $L_{2}$ regularisation term on entity and relation representations to the training objective.
Lastly, \citet{DBLP:conf/icml/LacroixUO18} observed systematic improvements by replacing the $L_{2}$ norm with a nuclear tensor 3-norm. 

\subsection{Item recommendation} \label{ssec:item}

\subsubsection{The implicit feedback scenario} In a recommendation system, let $u \in \mathcal{U}$ and $i \in \mathcal{I}$ be a user and a item, respectively, with $|\mathcal{U}| = N$ and $|\mathcal{I}| = M$. Then, let $\mathbf{X} \in \mathbb{R}^{N \times M}$ be the user-item interaction matrix, where the entry $x_{ui}$ is equal to $1$ if user $u$ interacted with item $i$, 0 otherwise. In the following, we consider the \textit{implicit} feedback scenario, meaning that the existence of a user-item interaction (i.e., $x_{ui} = 1$) does not necessarily mean that the user likes the item, but it only means that the user \textit{interacted with} the item (e.g., click, view, purchase). Likewise, when $x_{ui} = 0$, we are not assuming that the user does not like the item, but we are stating that no interaction has occurred between the user and the item yet. In this respect, the item recommendation problem is the task of predicting the score for the 0 entries in the user-item interaction matrix. 

\subsubsection{User-item score predictors} A \textit{user-item score predictor} is a differentiable model where users in $\mathcal{U}$ and items in $\mathcal{I}$ are mapped to a continuous embedding space, and the likelihood of their interaction is a function of their representations. That is, we define a parametric \emph{scoring function} $\varrho_\theta: \mathcal{U} \times \mathcal{I} \mapsto \mathbb{R}$ that, given a user-item pair $(u, i)$, estimates the likelihood that user $u$ may interact with item $i$ in the future.

\subsubsection{Scoring functions}
In item recommendation, collaborative filtering~\cite{DBLP:journals/fthci/EkstrandRK11} (CF) has long dominated as the leading paradigm, suggesting that similar users may likely interact with similar items. Among the pioneer approaches in CF, nearest-neighbour ones leverage the concept of users' or items' similarities based upon some heuristics. For instance, ItemkNN~\cite{DBLP:conf/www/SarwarKKR01} predicts whether a user $u$ and an item $i$ could interact depending on the similarity between $i$ and the other items $u$ has interacted with:
\begin{equation}
    \varrho^{\text{ItemkNN}}(u, i) = \sum_{j \in \mathcal{I}_u^{+}} \sigma(i, j),
\end{equation}
where $\mathcal{I}_u^{+}$ is the set of items interacted by user $u$, while $\sigma(\cdot)$ is a similarity function (e.g., the cosine similarity) computed between the $i$-th and $j$-th column vectors of the user-item interaction matrix. For computational purposes, often times the set $\mathcal{I}_u^{+}$ is restricted to the $k$-most similar items. The dual approach, UserkNN~\cite{DBLP:conf/cscw/ResnickISBR94}, estimates the interaction score of $(u, i)$ based upon the similarity between $u$ and the other users $i$ has been interacted by: 
\begin{equation}
    \varrho^{\text{UserkNN}}(u, i) = \sum_{v \in \mathcal{U}_i^{+}} \sigma(u, v),
\end{equation}
where, complementarily to above, $\mathcal{U}_i^{+}$ is the set of users that interacted with item $i$. With the advent of machine learning, latent factor models have taken over CF-based recommendation. Specifically, we indicate with $\mathbf{u} \in \mathbb{R}^{d}$ and $\mathbf{i} \in \mathbb{R}^{d}$ the embeddings for user $u$ and item $i$ and, generally, $d \ll N, M$. One of the widely-popular approaches, matrix factorisation~\cite{DBLP:journals/computer/KorenBV09} (MF), estimates the user-item interaction score through the dot product of their embeddings:
\begin{equation}
    \varrho^{\text{MF}}(u, i) = \mathbf{u}^{\top} \mathbf{i} = \sum_{f=1}^{d} \mathbf{u}_f\mathbf{i}_f.
\end{equation}
By leveraging the representational power of deep neural networks, in neural collaborative filtering~\cite{DBLP:conf/www/HeLZNHC17} (NCF), the authors propose to predict the user-item interaction score through the following:
\begin{equation}
    \varrho^{\text{NCF}}(u, i) = \sigma(\mathbf{W} \cdot \left(\Psi([\mathbf{u}, \mathbf{i}]) + \mathbf{u} \odot \mathbf{i})\right),
\end{equation}
\noindent where $\sigma(\cdot)$ is an activation function (\eg the sigmoid), $\mathbf{W}$ is a weight matrix, $\Psi(\cdot)$ is a multilayer perceptron where the input is the concatenation of the user and item embeddings, and $\odot$ is the element-wise product. In SimpleX~\cite{DBLP:conf/cikm/MaoZWDDXH21}, the authors introduce a score function which is based upon the importance of the items interacted by each user:
\begin{equation}
    \varrho^{\text{SimpleX}}(u, i) = \sigma(\alpha \mathbf{u} + (1 - \alpha)\mathbf{W}\cdot\mathbf{u}'),
\end{equation}
where $\sigma(\cdot)$ is a similarity function (\eg the cosine similarity), $\alpha$ is a weighting hyper-parameter, $\mathbf{W}$ is a weight matrix, and $\mathbf{u}'$ is obtained as the weighted aggregation (e.g., through attention) of the embeddings of the items interacted by $u$.

\subsubsection{Training objectives}

As recognised in the related literature, approaches in item recommendation mainly adopt five \textit{training objectives}. Indeed, the most popular one is Bayesian personalised ranking (BPR), whose rationale drives from the concept of interacted (i.e., positive) and non-interacted (i.e., negative) items for each user in the catalogue. Let $\mathcal{T} = \{(u, i, j) \; | \; i \in \mathcal{I}_u^+ \land j \in \mathcal{I} \setminus \mathcal{I}_u^{+}\}$ be a set of triples, where $\mathcal{I}_u^{+}$ is the set of all positive items for user $u$ and $\mathcal{I} \setminus \mathcal{I}_u^{+}$ is the set of all negative items for user $u$. BPR seeks to maximise the posterior probability of each user $u$ preferring a positive item $i$ over a negative item $j$. Thus, the BPR loss function is calculated as:
\begin{equation}
    \mathcal{L}(\Theta) = - \sum_{(u, i, j) \in \mathcal{T}} \text{ln } \sigma(\hat{x}_{ui} - \hat{x}_{uj}),
\end{equation}
where $\Theta$ is the vector of all model's weights (e.g., in the case of MF, the user and item embeddings), $\sigma(\cdot)$ is the sigmoid function, and $\hat{x}_{ui}, \hat{x}_{uj}$ are the predicted scores for the pairs of user $u$ with the positive item $i$, and user $u$ with the negative item $j$. 

The pairwise hinge loss (PH) works by maximising the distance between the user-positive item pair and the user-negative item pair, at least under a certain marginal threshold. Let $\mathcal{T}^+ = \{(u, i) \; | \; i \in \mathcal{I}_u^{+} \}$ and $\mathcal{T}^- = \{(u, j) \; | \; j \in \mathcal{I} \setminus \mathcal{I}_u^{+}\}$ be the sets of pairs of user and their positive items, and user and their negative items. Then, the PH loss is defined as:
\begin{equation}
    \mathcal{L}(\Theta) = \sum_{(u,i) \in \mathcal{T}^{+}} \sum_{(u,j) \in \mathcal{T}^{-}} w_{ui} [m + ||\mathbf{e}_u - \mathbf{e}_i||^2 - ||\mathbf{e}_u - \mathbf{e}_j||^2]_+,
\end{equation}
where $w_{ui}$ is a ranking loss weight, $m > 0$ is the margin size, and $[x]_+ = max(x, 0)$. 

The binary cross-entropy loss (BCE) is adopted for the task of binary classification, discerning whether an item should be classified as positive or negative for a given user:

\begin{equation}
    \mathcal{L}(\Theta) = - \sum_{(u, k) \in \mathcal{T}^{+} \cup \mathcal{T}^{-}} x_{uk} \text{ln}(\hat{x}_{uk}) + (1 - x_{uk}) \text{ln}(1 - \hat{x}_{uk}),
\end{equation}
where $x_{uk} = 1$ if $k \in \mathcal{I}_u^{+}$, 0 otherwise. The multiclass version of the previous loss implies the recast of the item recommendation problem to a multiclass classification one where, for each user, the model predicts which of the items (from the whole catalogue) the user will likely interact with. Such a loss function is named softmax cross-entropy loss (SCE), and is calculated as:
\begin{equation}
    \mathcal{L}(\Theta) = - \sum_{u \in \mathcal{U}} \sum_{k \in \mathcal{I}} x_{uk} \text{ln}\left(\frac{\text{exp}(\hat{x}_{uk})}{\sum_{v \in \mathcal{I}} \text{exp}(\hat{x}_{uv})}\right),
\end{equation}
where $\text{exp}(\cdot)$ is the exponential function.

The mean square error (MSE) measures the distance between the true and predicted score for each user-item pair:
\begin{equation}
    \mathcal{L}(\Theta) = \sum_{(u, k) \in \mathcal{T}^+ \cup \mathcal{T}^-} (x_{uk} - \hat{x}_{uk})^2.
\end{equation}

While we defined $\mathcal{I} \setminus \mathcal{I}_u^{+}$ as the set of \textit{all} negative items for user $u$, oftentimes it is computationally-expensive to work with such a large set of elements. 
For this reason, the common approach is to sample a subset of negative items for each user, namely, \textit{negative sampling}. Indeed, the authors of BPR suggest to sample, for each pair of user-positive item, only one negative item for that user. Nevertheless, the solution can be trivially generalised to sampling $N^{-}$ negative items for each user. 

Let $\mathcal{I}_u^{-}$ be the set of sampled negative items for user $u$, with $|\mathcal{I}_u^{-}| = N^{-}$. An additional training objective which have been successfully adopted over the last few years in the literature is the contrastive loss. For instance, the authors in SimpleX propose a cosine contrastive loss (CC) defined as:
\begin{equation}
    \mathcal{L}(\Theta) = \sum_{(u, i) \in \mathcal{T}^{+}} (1 - \hat{x}_{ui}) + \frac{w}{N^{-}} \sum_{j \in \mathcal{I}_u^{-}} [\hat{x}_{uj} - m]_{+},
\end{equation}
where $m$ is a margin size as in the PH loss, while $w$ controls the relative weights of the two addendums of the loss.

\subsubsection{Regularisers}
Similarly to the outlined models for link prediction, also most recommendation systems leverage $L_2$ regularisation on the user and item embeddings to control their magnitude, thus preventing overfitting~\cite{DBLP:conf/nips/SalakhutdinovM07, DBLP:journals/computer/KorenBV09}. Besides that, no other regularisation techniques are generally explored in the literature. Hence, our work may also serve as a preliminary analysis to understand whether and to what extent different regularisation techniques from the link prediction domain can impact on the recommendation performance.

%% file: src/proposed.tex

%
\section{Item Recommendation as Link Prediction}
\label{sec:cast}
Note that item recommendation models, outlined in \cref{ssec:item}, can be cast as a particular case of link prediction in Knowledge Graphs.
%
%
More specifically, user-item score predictors $\varrho_\theta: \mathcal{U} \times \mathcal{I} \mapsto \mathbb{R}$ can be seen as learning a ranking between missing triples in a Knowledge Graph $\mathcal{G}$, where the set of entities corresponds to the union of the sets of users and items $\mathcal{E} = \mathcal{U} \cup \mathcal{I}$, the set of relations corresponds to a single relation $\mathcal{R} = \{ \text{interactsWith} \}$, and the graph $\mathcal{G}$ to complete is given by the observable interactions between users and items:
\begin{equation}
\mathcal{G} = \{ \langle u, \text{interactsWith}, i \rangle \mid x_{ui} = 1 \}.
\end{equation}
This enables the off-the-shelf application of state-of-the-art neural link prediction methods to the item recommendation problem.
In this work, we begin to investigate whether off-the-shelf neural link prediction models can be used for item recommendation.
More formally, we aim to answer the following research questions.
\subsection{Research questions}

\begin{description}
    \item[\textbf{RQ1}] To what extent can out-of-the-box link prediction models produce accurate results compared to state-of-the-art recommendation systems on the item recommendation task?
    \item[\textbf{RQ2}] What is the influence of hyper-parameters to make link prediction models generalise well on the recommendation tasks?
\end{description}
%
%

%
%

%% file: src/experiments.tex
\section{Experimental study}
\label{sec:experiments}
This section delves into the experimental settings of this paper. First, we describe the recommendation datasets along with their statistics. Second, we summarise the main technical aspects of the selected recommendation and link prediction baselines. Then, we indicate how we adapted the evaluation of link prediction models to fit the item recommendation task. Finally, we provide reproducibility details for our analysis.

\subsection{Datasets}

We use three widely-adopted recommendation datasets, namely, Gowalla, Yelp 2018, and Amazon Book (\Cref{tab:datasets}). 

\noindent $\bullet$ \textbf{Gowalla}~\cite{DBLP:conf/www/LiangCMB16} records the check-in history of users, where each check-in event corresponds to a location. It counts 29,858 users and 40,981 items, with a total of 1,027,370 recorded interactions (i.e., the sparsity is 0.9992). 

\noindent $\bullet$ \textbf{Yelp 2018}~\cite{DBLP:conf/sigir/0001DWLZ020} collects users' reviews about local business and derives from the 2018 Yelp challenge. The number of users and items is 31,668 and 38,048, respectively, while the number of user-item interactions amounts to 1,561,406 (i.e., the sparsity is 0.9987). 

\noindent $\bullet$ \textbf{Amazon Book}~\cite{DBLP:conf/www/HeM16} is a product category from the popular Amazon recommendation dataset. It collects the clicks metadata accounting for 52,643 users and 91,599 items, with a total number of 2,984,108 interactions (i.e., the sparsity is 0.9994). 

\input{tables/datasets}

\subsection{Models}

\subsubsection{Item recommendation} We select ten approaches for item recommendation and summarise their main strategies.

\noindent $\bullet$ \textbf{MostPop} is an unpersonalised recommendation method based on the popularity of items in the catalog.

\noindent $\bullet$ \textbf{Random} is an unpersonalised recommendation method that retrieves random item lists for each user.

\noindent $\bullet$ \textbf{UserkNN}~\cite{DBLP:conf/cscw/ResnickISBR94} is a neighbor-based model which recommends new items based upon similarities between users' profiles (i.e., their interaction history).

\noindent $\bullet$ \textbf{ItemkNN}~\cite{DBLP:conf/www/SarwarKKR01} follows the same rationale as UserkNN but from the item's perspective.



\noindent $\bullet$ \textbf{NGCF}~\cite{DBLP:conf/sigir/Wang0WFC19} is among the pioneer approaches exploiting graph neural networks (GNNs) for item recommendation. Its message-passing schema consists of the aggregation the neighbourhood information and the inter-dependencies among the \textit{ego} and the \textit{neighbourhood} nodes (calculated as the Hadamard product of the two embeddings). Through the message-passing performed for \textit{L}-hops, the refined user and item embeddings are used to perform the common BPRMF score and loss functions.

\noindent $\bullet$ \textbf{DGCF}~\cite{DBLP:conf/sigir/WangJZ0XC20} is a graph-based recommendation systems assuming that user-item interactions can be disentangled into independent intents, where each describes one aspect of the user's preference towards the item. The model performs graph structure learning to refining the adjacency matrix driven from the recognised and learned intents. As in NGCF, the score function is based upon MF, while the loss function introduces an additional term to the BPR loss to encourage the independence of the intents.  

\noindent $\bullet$ \textbf{LightGCN}~\cite{DBLP:conf/sigir/0001DWLZ020} is a graph-based recommender system proposing a light-weight version of the graph convolutional network (GCN) layer to provide superior accuracy performance. In terms of architectural choices, the model removes feature transformations and non-linearities during the message-passing with respect to the GCN layer. The score and loss functions follow the BPRMF schema.

\noindent $\bullet$ \textbf{SGL}~\cite{DBLP:conf/sigir/WuWF0CLX21} applies the concepts of self-supervised and contrastive learning to graph-based recommendation. Based upon the LightGCN layer, the model performs node/edge dropout and random walk operations to build different versions of the adjacency matrix, thus learning also different node views. The self-supervised contrastive loss (added to the BPR component) encourages the consistency among different views of the same node and the divergence among different nodes. As for the previous graph-based approaches, MF is used as score function.

\noindent $\bullet$ \textbf{UltraGCN}~\cite{DBLP:conf/cikm/MaoZXLWH21} is a graph-based recommendation system which introduces a novel message-passing schema approximating the infinite-layer propagation through a single (simplified) node update iteration. The adjacency matrix is normalised by considering the asymmetric weighting of connected nodes in user-user and item-item connections. In place of BPR, two loss components are introduced to tackle the over-smoothing effect and learn from item-item relations. As usual, the score function is based upon MF.

\noindent $\bullet$ \textbf{GFCF}~\cite{DBLP:conf/cikm/ShenWZSZLL21} is a memory-based approach which does not use any trainable model weight. The approach exploits graph signal processing theory to demonstrate that existing approaches in item recommendation may fall into one unified framework leveraging graph convolution. Thus, the authors design a strong closed-form algorithm outperforming most of the trainable approaches previously introduced in the literature.

\subsubsection{Link prediction} We select three state-of-the-art approaches for link prediction. We summarize their main strategies in the following.

\noindent $\bullet$ \textbf{DistMult}~\cite{DBLP:journals/corr/YangYHGD14a} is a decomposition technique for link prediction in which entities and predicates are combined through a tri-linear dot product to predict the interaction score for the triple. Each entity has only one representation, independent on it appearing as a subject or an object in the triples.

\noindent $\bullet$ \textbf{CP}~\cite{DBLP:conf/icml/LacroixUO18} is similar to DistMult, where the only difference is that each entity comes with a double representation, one for when it appears as a subject and the other for the object. 

\noindent $\bullet$ \textbf{ComplEx}~\cite{DBLP:conf/icml/TrouillonWRGB16} is similar to the previous link prediction approaches, where the main differences lay in the representation of entities and predicates through complex vectors. For this reason, the scoring function is also modified to the usual tri-linear dot product, as it works on the complex representation of the subject and predicate, the complex conjugate of the object, and finally takes the real part of the predicted score.

\subsection{Evaluation}

While the selected recommender systems were tuned and evaluated through the framework Elliot~\cite{DBLP:conf/recsys/AnelliMPBSN23, DBLP:conf/sigir/AnelliBFMMPDN21, DBLP:conf/um/MalitestaPANF23}, the link prediction models were tuned and evaluated with the popular framework LibKGE for knowledge graph embeddings~\cite{Teach2020YouCT}. However, as the latter is designed to address the task of link prediction which, as stated, is a generalization of the recommendation task, modifications were needed to make the link prediction results comparable to those of item recommendation. Apart from a re-casting of the recommendation dataset into a knowledge graph one (users and items are entities connected through one single interaction type, <<interactsWith>>), the calculation of the Recall@$k$ metric was added to LibKGE, as its formulation in recommender systems is different from that utilized in link prediction-alike tasks. As for the nDCG@\textit{k}, LibKGE was modified in a way the recommendation lists for each user were stored after the training of each model. Thus, Elliot was eventually exploited to compute the nDCG@$k$.

\input{tables/results}

\subsection{Reproducibility}

We used the same train/test splitting adopted for the training and test of the recommender systems~\cite{DBLP:conf/recsys/AnelliMPBSN23, DBLP:conf/um/MalitestaPANF23}. However, to perform hyper-parameter tuning of the link prediction approaches, we retained the 10\% of the training set as the validation set and used the Recall@20 as the validation metric. Three different validation splittings were considered for the sake of generalization. Regarding the explored hyper-parameters, and similarly to what was done in~\cite{Teach2020YouCT}, we tested three training strategies (i.e., 1vsAll, KvsAll, and negative sampling), two losses (i.e., BCE and KL), two batch sizes (i.e., 1024 and 2048), two optimisers (i.e., Adam and Adagrad), four learning rates (i.e., 0.0001, 0.001, 0.01, 0.1), two regularisers (i.e., N3, LP), and six regularisation weights (i.e., 1e-06, 1e-05, 1e-04, 1e-03, 1e-02, 1e-01). The embedding size was kept fixed at 64 to be consistent with the settings followed for the recommender systems.

%% file: tables/datasets.tex
\begin{table}[!t]
\centering
\caption{Statistics of the tested datasets.}
\label{tab:datasets}
\begin{tabular}{lrrrr}
\toprule
\textbf{Datasets} & \textbf{Users} & \textbf{Items} & \textbf{Interactions} & \textbf{Sparsity} \\
\cmidrule{1-5}
Gowalla & 29,858 & 40,981 & 1,027,370 & 0.9992 \\
Yelp 2018 & 31,668 & 38,048 & 1,561,406 & 0.9987 \\
Amazon Book & 52,643 & 91,599 & 2,984,108 & 0.9994 \\
\bottomrule
\end{tabular}
\end{table}

%% file: tables/results.tex
\begin{table*}[!t]
\centering
\caption{Performance comparison of different families of recommender systems (i.e., \textit{Reference}, \textit{Item Recommendation}, and \textit{Link Prediction}) on Gowalla, Yelp 2018, and Amazon Book, as Recall@20 and nDCG@20. The top-\textit{n} performing models (\textit{n} = 5) are reported in boldface and marked with (\textit{n}) on the right-hand side of the metric value indicating its relative position.}
\label{tab:baselines}
\begin{tabular}{llcccccc}
\toprule
\textbf{Families} & \textbf{Models} & \multicolumn{2}{c}{\textbf{Gowalla}} & \multicolumn{2}{c}{\textbf{Yelp 2018}} & \multicolumn{2}{c}{\textbf{Amazon Book}} \\ \cmidrule{3-8} 
\multicolumn{2}{c}{} & \multicolumn{1}{c}{Recall@20} & \multicolumn{1}{c}{nDCG@20} & \multicolumn{1}{c}{Recall@20} & \multicolumn{1}{c}{nDCG@20} & \multicolumn{1}{c}{Recall@20} & \multicolumn{1}{c}{nDCG@20} \\ \cmidrule{1-8} 
\multirow{2}{*}{\textit{Reference}~\cite{DBLP:conf/recsys/AnelliMPBSN23}} & MostPop & 0.0416 & 0.0316 & 0.0125 & 0.0101 & 0.0051 & 0.0044 \\
& Random & 0.0005 & 0.0003 & 0.0005 & 0.0004 & 0.0002 & 0.0002 \\
\cmidrule{1-8} 
\multirow{8}{*}{\textit{Item Rec.}~\cite{DBLP:conf/recsys/AnelliMPBSN23}} & UserkNN & \textbf{0.1685 (5)} & \textbf{0.1370 (5)} & \textbf{0.0630 (5)} & \textbf{0.0528 (5)} & \textbf{0.0582 (4)} & \textbf{0.0477 (4)} \\
& ItemkNN & 0.1409 & 0.1165 & 0.0610 & 0.0507 & \textbf{0.0634 (3)} & \textbf{0.0524 (3)} \\
& NGCF & 0.1556 & 0.1320 & 0.0556 & 0.0452 & 0.0319 & 0.0246 \\
& DGCF & \textbf{0.1736 (4)} & \textbf{0.1477 (4)} & 0.0621 & 0.0505 & 0.0384 & 0.0295 \\
& LightGCN & \textbf{0.1826 (3)} & \textbf{0.1545 (2)} & 0.0629 & 0.0516 & 0.0419 & 0.0323 \\
& SGL & \multicolumn{1}{c}{---} & --- & \textbf{0.0669 (3)} & \textbf{0.0552 (3)} & 0.0474 & 0.0372 \\
& UltraGCN & \textbf{0.1863 (1)} & \textbf{0.1580 (1)} & \textbf{0.0672 (2)} & \textbf{0.0553 (2)} & \textbf{0.0688 (2)} & \textbf{0.0561 (2)} \\
& GFCF & \textbf{0.1849 (2)} & \textbf{0.1518 (3)} & \textbf{0.0697 (1)} & \textbf{0.0571 (1)} & \textbf{0.0710 (1)} & \textbf{0.0584 (1)} \\
\cmidrule{1-8} 
\multirow{3}{*}{\textit{Link Pred.}} 
& DistMult & 0.1477 & 0.1182 & 0.0465 & 0.0375 & 0.0426 & 0.0330 \\
& CP & 0.1420
& 0.1076 & 0.0367 & 0.0293 & 0.0400 & 0.0309 \\
& ComplEx & 0.1619 & 0.1263 & \textbf{0.0662 (4)} & \textbf{0.0539 (4)} & \textbf{0.0492 (5)} & \textbf{0.0382 (5)} \\
\bottomrule
\end{tabular}
\end{table*}

%% file: src/results_discussion.tex
\section{Results and discussion}
\label{sec:results}

This section presents the preliminary results of our analysis, by addressing the two research questions of our work: \textbf{RQ1)} To what extent can out-of-the-box link prediction models produce accurate results compared to state-of-the-art recommendation systems on the item recommendation task? \textbf{RQ2)} What is the influence of hyper-parameters to make link prediction models generalise well on the recommendation tasks?

\subsection{RQ1. Link prediction \textit{vs.} item recommendation models}

Initial experiments were conducted to test the efficacy of state-of-the-art link prediction approaches (i.e., DistMult, CP, and ComplEx) when trained and tested for the task of item recommendation on the three selected recommendation datasets (i.e., Gowalla, Yelp 2018, and Amazon Book). In this respect, the link prediction models were compared against popular recommendation systems (i.e., MostPop, Random, UserkNN, ItemkNN, NGCF, DGCF, LightGCN, SGL, UltraGCN, GFCF), whose results have been directly picked from previous works that share the same experimental settings~\cite{DBLP:conf/recsys/AnelliMPBSN23, DBLP:conf/um/MalitestaPANF23}.

Results, in terms of Recall@20, and nDCG@20, are reported in~\Cref{tab:baselines}. Noticeably, on Yelp 2018 and Amazon Book, the three link prediction models always place in the top-5 most accurate models, outperforming most traditional and GNNs-based recommendation models. On Gowalla, even not being in the top-5 performing models, they still succeed in outperforming strong recommendation baselines such as ItemkNN and NGCF. Thus, our assumption that link prediction approaches built upon factorisation and the dot product of the entities/predicates embeddings can work quite well on the item recommendation task is also empirically validated. Once again, we recall that no specific architectural modification was done to the three link prediction approaches, which can work completely out-of-the-box the item recommendation problem. 

\subsection{RQ2. Hyper-parameter analysis of link prediction approaches}

To complement the findings from these initial experiments, we also conducted a hyper-parameter analysis on the impact of embedding size on the final recommendation performance for the three link prediction models. Specifically, we explore the performance of such models for the embedding size in the range [128, 256, 512, 1024], while we recall that the default value (for the previous results) is 64. Results, in terms of Recall@20 for all three tested datasets, are reported in~\Cref{fig:ablation}. As evident from the plots, and confirming the outcomes from the related literature, the performance of link prediction models on the item recommendation task generally increases monotonically along with the embedding size. Among the three selected models, the analysed trend is more observable for ComplEx.

\input{figures/ablation}

%% file: figures/ablation.tex
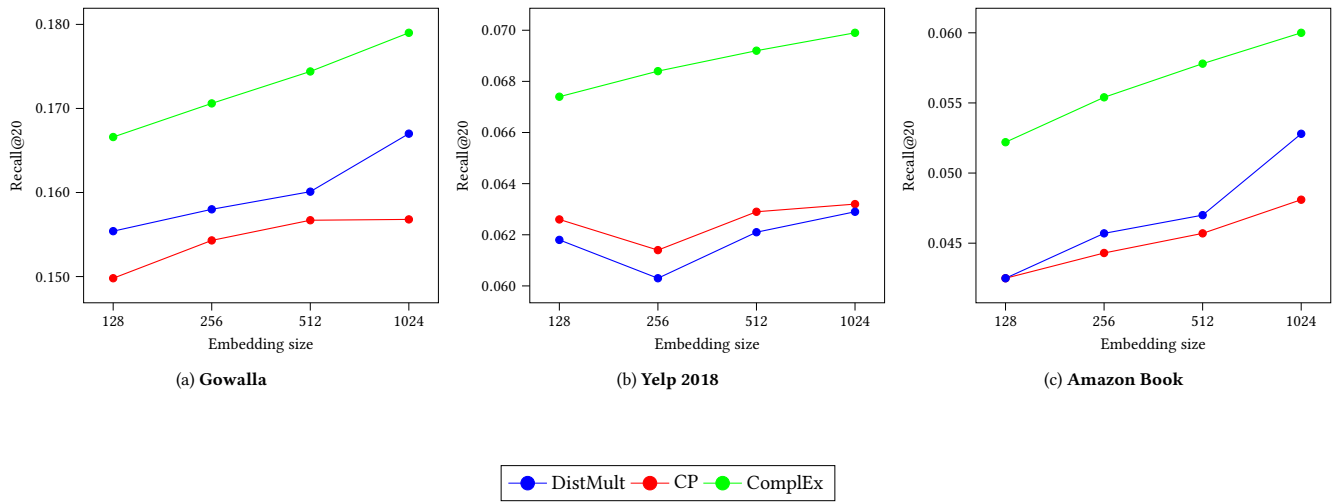
\begin{figure*}[!t]
\captionsetup[subfigure]{font=large,labelfont=large}
\centering
\begin{adjustbox}
{width=\textwidth,center}
\subfloat[Gowalla]{
    \input{figures/gowalla}
}
\subfloat[Yelp 2018]{
    \input{figures/yelp}
}
\subfloat[Amazon Book]{
    \input{figures/amazon}
}
\end{adjustbox}

\vspace{3em}

\begin{adjustbox}{width=0.25\textwidth,center}
\begin{tikzpicture}
        \begin{customlegend}[legend columns=3,
        legend entries={DistMult, CP, ComplEx}]
        \addlegendimage{blue, mark=*, mark size=3pt}
        \addlegendimage{red, mark=*, mark size=3pt}
        \addlegendimage{green, mark=*, mark size=3pt}
        \end{customlegend}
    \end{tikzpicture}
\end{adjustbox}
\caption{Hyper-parameter analysis on the influence of embedding size for DistMult, CP, and ComplEx, on the three tested recommendation datasets. Results refer to Recall@20.}
\label{fig:ablation}
\end{figure*}

%% file: figures/gowalla.tex
\begin{tikzpicture}
  \begin{axis}[
    yticklabel style={
        /pgf/number format/.cd,
        fixed, fixed zerofill,
        precision=3
    },
    scaled y ticks=false,
    xlabel={Embedding size},
    ylabel={Recall@20},
    tick align=outside,
    tick pos=left,
    xmin=0.7, xmax=4.3,
    x grid style={darkgray!50},
    xtick style={color=black},
    xticklabels={128, 256, 512, 1024},
    xtick={1, 2, 3, 4},
    y grid style={darkgray!50},
    ytick style={color=black}
  ]
    \addplot [semithick, red, forget plot, mark=*] coordinates {
    (1,0.1498)
    (2,0.1543)
    (3,0.1567)
    (4,0.1568)
    };
    \addplot [semithick, blue, forget plot, mark=*] coordinates {
    (1,0.1554)
    (2,0.1580)
    (3,0.1601)
    (4,0.1670)
    };
    \addplot [semithick, green, forget plot, mark=*] coordinates {
    (1,0.1666)
    (2,0.1706)
    (3,0.1744)
    (4,0.1790)
    };
\end{axis}
\end{tikzpicture}

%% file: figures/yelp.tex
\begin{tikzpicture}
  \begin{axis}[
    yticklabel style={
        /pgf/number format/.cd,
        fixed, fixed zerofill,
        precision=3
    },
    scaled y ticks=false,
    xlabel={Embedding size},
    ylabel={Recall@20},
    tick align=outside,
    tick pos=left,
    xmin=0.7, xmax=4.3,
    x grid style={darkgray!50},
    xtick style={color=black},
    xticklabels={128, 256, 512, 1024},
    xtick={1, 2, 3, 4},
    y grid style={darkgray!50},
    ytick style={color=black}
  ]
    \addplot [semithick, red, forget plot, mark=*] coordinates {
    (1,0.0626)
    (2,0.0614)
    (3,0.0629)
    (4,0.0632)
    };
    \addplot [semithick, blue, forget plot, mark=*] coordinates {
    (1,0.0618)
    (2,0.0603)
    (3,0.0621)
    (4,0.0629)
    };
    \addplot [semithick, green, forget plot, mark=*] coordinates {
    (1,0.0674)
    (2,0.0684)
    (3,0.0692)
    (4,0.0699)
    };
    
\end{axis}
\end{tikzpicture}

%% file: figures/amazon.tex
\begin{tikzpicture}
  \begin{axis}[
    yticklabel style={
        /pgf/number format/.cd,
        fixed, fixed zerofill,
        precision=3
    },
    scaled y ticks=false,
    xlabel={Embedding size},
    ylabel={Recall@20},
    tick align=outside,
    tick pos=left,
    xmin=0.7, xmax=4.3,
    x grid style={darkgray!50},
    xtick style={color=black},
    xticklabels={128, 256, 512, 1024},
    xtick={1, 2, 3, 4},
    y grid style={darkgray!50},
    ytick style={color=black}
  ]
    \addplot [semithick, red, forget plot, mark=*] coordinates {
    (1,0.0425)
    (2,0.0443)
    (3,0.0457)
    (4,0.0481)
    };
    \addplot [semithick, blue, forget plot, mark=*] coordinates {
    (1,0.0425)
    (2,0.0457)
    (3,0.0470)
    (4,0.0528)
    };
    \addplot [semithick, green, forget plot, mark=*] coordinates {
    (1,0.0522)
    (2,0.0554)
    (3,0.0578)
    (4,0.0600)
    };
\end{axis}
\end{tikzpicture}

%% file: src/conclusion.tex
\section{Conclusion and Future Work}
\label{sec:conclusions}

This work presented a preliminary attempt to bridge the gap between two similar (but generally regarded as distinct) tasks: link prediction and item recommendation. Assuming the models for the two tasks are largely built upon the factorisation of the entities (user-item) interaction matrix, and their score is usually computed through the dot product of the learned embeddings, we started by theoretically showing the two model families have several aspects in common. That is, item recommendation may be re-casted as a link prediction task, where users and items are entities in a knowledge graph with a single-type predicate: <<interactsWith>>. To empirically validate our initial assumptions, we also run preliminary experiments, by taking three out-of-the-box link prediction models (i.e., DistMult, CP, and ComplEx), and training/testing them on the item recommendation task. Specifically, we evaluated their accuracy recommendation performance on three popular recommendation datasets (i.e., Gowalla, Yelp 2018, and Amazon Book), against ten state-of-the-art recommendation approaches (i.e., MostPop, Random, UserkNN, ItemkNN, NGCF, DGCF, LightGCN, SGL, UltraGCN, and GFCF). Overall, results proved that link prediction models can perform adequately well on the item recommendation task, even without specific modifications to their architectures. Finally, we noticed that higher values of the embedding size of such models can lead to improved recommendation results. This opens new paths to explore in the future. We plan to explore if and to what extent other hyper-parameters of the link prediction models may affect their performance on the recommendation task, such as the adoption of bi-directional predicates to indicate the <<interactsWith>> relation, or other training paradigms.